# Emerging Immersive Communication Systems: Overview, Taxonomy, and Good Practices for QoE Assessment

Pablo Pérez[1]*, Ester Gonzalez-Sosa[1], Jesús Gutiérrez[2] and Narciso García[2]

[1]eXtended Reality Lab, Nokia, Madrid, Spain, [2]Grupo de Tratamiento de Imágenes, Universidad Politécnica de Madrid, Madrid, Spain

Several technological and scientific advances have been achieved recently in the fields of immersive systems (e.g., 360-degree/multiview video systems, augmented/mixed/virtual reality systems, immersive audio-haptic systems, etc.), which are offering new possibilities to applications and services in different communication domains, such as entertainment, virtual conferencing, working meetings, social relations, healthcare, and industry. Users of these immersive technologies can explore and experience the stimuli in a more interactive and personalized way than previous technologies (e.g., 2D video). Thus, considering the new technological challenges related to these systems and the new perceptual dimensions and interaction behaviors involved, a deep understanding of the users' Quality of Experience (QoE) is required to satisfy their demands and expectations. In this sense, it is essential to foster the research on evaluating the QoE of immersive communication systems, since this will provide useful outcomes to optimize them and to identify the factors that can deteriorate the user experience. With this aim, subjective tests are usually performed following standard methodologies (e.g., ITU recommendations), which are designed for specific technologies and services. Although numerous user studies have been already published, there are no recommendations or standards that define common testing methodologies to be applied to evaluate immersive communication systems, such as those developed for images and video. Taking this into account, a revision of the QoE evaluation methods designed for previous technologies is required to develop robust and reliable methodologies for immersive communication systems. Thus, the objective of this paper is to provide an overview of existing immersive communication systems and related user studies, which can help on the definition of basic guidelines and testing methodologies to be used when performing user tests of immersive communication systems, such as 360-degree video-based telepresence, avatar-based social VR, cooperative AR, etc.

# 1 INTRODUCTION

In the past years, and even more since the COVID-19 pandemic, immersive communication systems (ICS) have gained a lot of interest yet beginning to be considered as the new form of computer-mediated communications. This is mainly due to the technological advances achieved in 360-degree/multiview video systems, augmented/mixed/extended/virtual reality systems (AR, MR, XR, VR), immersive audio-haptic systems, in combination with contributions from other relevant scientific areas such as computer vision, egocentric vision, spatial audio, mobile communications, volumetric capturing, etc. This new communication paradigm lays its foundation on the possibility of feeling immersed in a remote human-to-human communication, as a substantial change with respect to current videoconferencing-based systems.

From this perspective, immersive communication can be defined as *exchanging natural social signals with remote people, as in face-to-face meetings, and/or experiencing remote locations (and having the people there experience you) in ways that suspend disbelief in 'being there'* (Apostolopoulos et al., 2012). This *subjective experience of being in one place or environment, even when one is physically situated in another*, is called "sense of presence" (or simply "presence") (Witmer and Singer, 1998).

The number of extensive works done in ICS and other related fields have always motivated researchers towards finding a general framework to understand and describe all the literature from a common point of view. For instance, Rae et al. (2015) carried out a thorough revision for telepresence systems, whose aim is to elicit the sensation of "being there" *via* a facilitating or mediating technology. The hardware involved in such systems can go from a standard computer/laptop to more sophisticated ones such as cameras placed on robots. In this context, they identified seven dimensions to describe telepresence systems, namely: 1) how the telepresence is *initiated*, 2) the *physical environment* that will be used, 3) the degree of *mobility* that the system will offer, 4) the amount of *vision* that the system will provide to the remote user, 5) the *social environment*, e.g., relationships among stakeholders, 6) the ways of *communication* offered, and 7) the level of *independence* provided to users. Likewise, Anthony et al. (2019) conducted a survey on Collaborative MR, where VR goggles are the mainstream hardware device. There, previous related works were classified and described in terms of application areas and user interaction methodologies. To better enhance Collaborative MR, they suggested to focus on techniques for 1) annotating, 2) manipulating object cooperatively, and 3) reducing cognitive workload of task understanding while increase users' perceptual awareness.

In recent years, novel ICS have been proposed, exploring different paradigms from previous works. Examples can be found in prototypes or commercial systems as Microsoft Holoportation (Orts-Escolano et al., 2016), Google Starline (Lawrence et al., 2021), Mozilla Hubs, but also in many experimental solutions as VR Together (Gunkel et al., 2018), The Owl (Kachach et al., 2021), Shoulder of the Giant (Piumsomboon et al., 2019), Exleap (Izumihara et al., 2019), among many others. As a result, a new thorough revision of the literature is required. In particular, before focusing on the *use case* (telepresence, MR collaboration), we believe it is paramount to analyze the design and implementation of systems being proposed, how they have been built, and their corresponding challenges.

Apart from considering the new technological challenges related to these new emerging systems and their new perceptual dimensions and interaction behaviors, a deep understanding on how such systems can be evaluated from a subjective perspective of the users is required to satisfy their demands and expectations. This is addressed by the evaluation of the Quality of Experience (QoE), which was defined by Qualinet (Le Callet et al., 2012) [and standardised by the ITU in the recommendation P.10/G.100 (ITU-T, 2017)] as "the degree of delight or annoyance of the user with an application or service. It results from the fulfilment of his or her expectations concerning the utility and enjoyment of the application or service in the light of the user's personality and current state". With the recent development of ICS, it is essential to foster the research on evaluating the QoE of their users, which will help on identifying the factors that can deteriorate it and on providing useful outcomes to optimize them.

In this sense, subjective tests with users are fundamental, which are usually performed following standard methodologies (e.g., ITU recommendations), which are designed for specific technologies and services [e.g., for 2D video the recommendations ITU-R BT.500 (ITU-, 2019), ITU-T P.910 (ITU-, 2008), and ITU-T P.913 (ITU-, 2016a), for 3D video the recommendation ITU-T P.915 (ITU-, 2016b), etc.]. Although numerous user studies have been already published, there are still no recommendations or standards that define common testing methodologies to be applied to ICS, such as those developed for images and video. As already stated in Steed (2021), regarding the use of MR for human communications, *there is a real need to establish some common metrics that can be used to establish the quality of communication and interaction between people, standardized social tasks, and the quality of the MR-base communication through time*.

Ideally, a common vision would allow researchers to characterize any immersive communication system by their main perceptual and technological modules, as well as make fair comparisons between systems. Likewise, there exists no common/standard QoE assessment framework that could in turn assess the related technical and perceptual items of such systems. In this paper, we address these challenges with the following **contributions**:

- The proposal of a **new taxonomy** that allows to describe immersive communication systems based on four key perceptual features: visual communication (*Face*), remote presence (*Visit*), shared immersion (*Meet*), and embodied interaction (*Move*). We provide a comprehensive discussion of related immersive communication works taking into account the proposed taxonomy, showing that most existing systems can be classified into **seven different**

- **archetypes**, with similar perceptual and technical characteristics.
- A thorough analysis of the **QoE evaluation methods** carried out in these related works, to understand whether there are similarities in what concerns QoE methods between works that lie in the same space of the proposed taxonomy. We find that most QoE evaluation protocols follow a common structure: perform a set of communication tasks (deliberation, exploration, manipulation), under different system conditions (different systems, or different configurations of the same one), and observe its effect in several QoE features.
- The proposal of a list of **good practices** for QoE assessment in ICS, which could be the basis for future standardization work on the matter.

The article is organized around these contributions. **Section 2** presents the proposed taxonomy and uses it to describe related works of immersive communication systems. **Section 3** analyzes how QoE is assessed in such related works. **Section 4** provides a discussion on fundamental aspects of QoE assessment in immersive communication systems, as well as a set of good practices. Conclusions and an outlook on future research lines are provided in **Section 5**.

## 2 A TAXONOMY OF IMMERSIVE COMMUNICATION SYSTEMS

For the purpose of this article, we will define *immersive communication systems* as the ones which enable **immersive**, **remote**, and **synchronous** communication. As already mentioned in the introduction, **immersive** systems are characterized for creating the illusion of "being there" (a "sense of presence"), in a remote location and/or with remote people. This is done through several technical features: use of large or head-mounted displays, volumetric capture and representation of participants, transmission of immersive video from a remote area, etc. (Apostolopoulos et al., 2012).

As technology is evolving quickly, there are dozens of different proposals, mostly experimental, of ICS. Therefore, a taxonomy is needed to classify them according to their most relevant characteristics. A possible analysis line would be to detail the different technology blocks involved in each system, as proposed by Park and Kim (2022), or identify the most relevant use cases and their implementation details, as proposed by Wang et al. (2021). However, this easily results in a large number of classification elements, and therefore a much larger number of categories. Alternatively, Apostolopoulos et al. (2012) propose a single dimension analysis, describing immersive systems as a continuum: From *immersive communication* systems designed to support *natural conversation* between people, to *immersive collaboration* systems designed to support *sharing information* between people. Using a single dimension, however, makes it difficult for the framework to properly analyze all possible existing systems. Jonas et al. (2019) propose a taxonomy for social VR applications, based on three characteristics: *the self* (avatar), *interaction with others*, and *environment*. Schäfer et al. (2021) propose a similar one, also applicable to MR and AR systems. Both are, however, focused on AR/VR/MR systems, therefore being difficult to apply to other immersive technologies, such as light-field displays.

To build our taxonomy, we have studied the state of the art of ICS, analyzing their technology and proposed use cases, searching for similarities and differences. We have identified four fundamental elements which describe the basic *perceptual features* of the communication (**Section 2.1**), in a similar spirit as the taxonomies proposed by Apostolopoulos et al. (2012), Jonas et al. (2019), or Schäfer et al. (2021), but covering a broader range of immersive systems. These fundamental elements can be used to define seven archetypes (**Section 2.2**) which represent most of the existing immersive communication systems in the literature, as it will be shown in **Section 2.4**. Besides, we have identified the main technological components of the systems (**Section 2.3**), similarly to Park and Kim (2022) or Wang et al. (2021). We will show how our identified fundamental elements are tightly related with the technical components, in such a way that systems belonging to the same archetype use similar system components in their implementations (**Section 2.5**).

## 2.1 Fundamental Elements of Immersive Communication Systems

We assume that ICS try to emulate in-person communication, ideally aiming at achieving total equivalence to a physical meeting[1]. As this task is not achievable in general, different systems focus on different aspects of the communication. Based on this idea, we have identified four basic communication elements (depicted in **Figure 1**) that are the building blocks of those systems: *Face*, *Visit*, *Meet* and *Move*.

1. **Face** is the property of the system to transmit in real time a visual representation of the other person, e.g. through a video-conferencing system. This element enables **visual communication.** Seeing the other person is key to transmit non-verbal communication cues, including showing objects of the personal space.
2. **Visit** is the property of the system to transmit in real time a visual representation of the *surroundings* of the other person. This enables **remote presence**: seeing the physical environment of the other person and being able to operate and discuss about it.
3. **Meet** is the property of the system to represent the other person in the same (virtual or physical) space as the user. This enables **shared immersion**: Being immersed in the same (virtual or physical) environment and interacting with the same (virtual or physical) objects.

---

[1]It can be argued that replicating face-to-face communication is not necessarily the final goal of immersive systems (Hollan and Stornetta, 1992); however, *thinking about* the implications of having such equivalence has helped us to identify the fundamental building blocks described in this section.

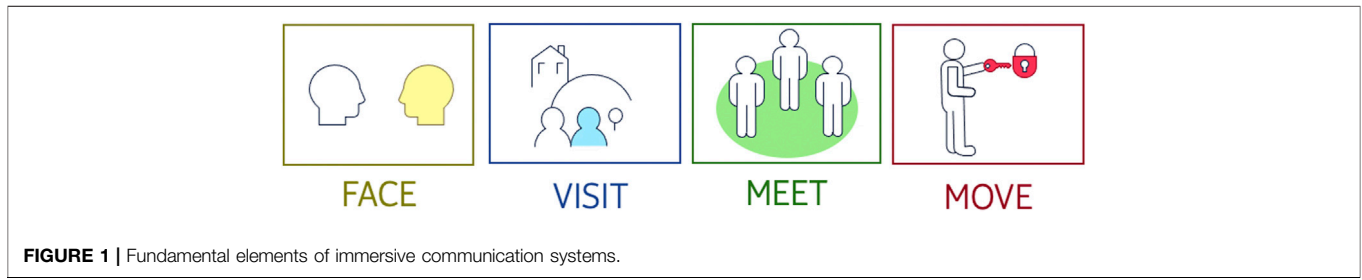

FIGURE 1 | Fundamental elements of immersive communication systems.

TABLE 1 | Fundamental elements of an immersive communication system.

| Element | Face | Visit | Meet | Move |
| --- | --- | --- | --- | --- |
| Perceptual prop | Visual communication | Remote presence | Shared immersion | Embodied interaction |
| Experience | I see you | I see what you see | I am with you | I control objects |
| Paradigm | Face-to-face | Shoulder-to-shoulder | Hand-in-hand | Hand-to-world |

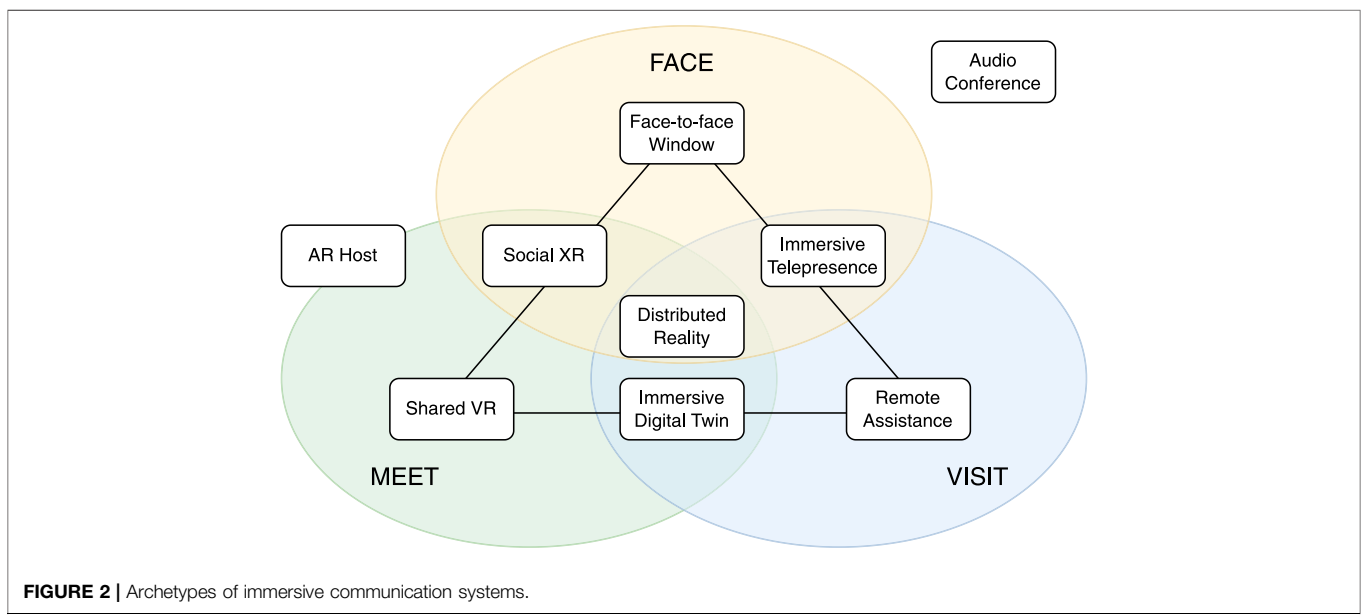

FIGURE 2 | Archetypes of immersive communication systems.

4. **Move** is the property of the system itself to represent the user within it and enable its **embodied interaction.** It means that the actions of the users are represented within the system and allow the user to interact with it.

Table 1 describes these elements from several perspectives: the main perceptual property provided by each element, as well as the communication *experience* (e.g., "I see you") and *paradigm* (e.g., "face-to-face") associated to each of them.

It is worth noting that these fundamental elements are described from the perspective of the *receiver* of the communication. This distinction is important because some communication systems are not symmetrical (Ens et al., 2019), and therefore they may provide different communication features to each side of the communication. E.g., transmitting the user environment in one direction (*Visit*) and the user image in the other (*Face*).

## 2.2 Archetypes

The first three fundamental elements (**Face**, **Meet**, and **Visit**) are related to human communication. They are not easy to fulfill simultaneously, and therefore they allow to classify systems depending on whether they focus on fulfilling one or the other. This allows us to define seven **archetypes** of immersive communication systems (plus two non-immersive). **Figure 2** shows graphically those archetypes and their relations.

Three archetypes refer to communication systems focusing on only one of the fundamental elements. They are the simplest ones, and in fact, several commercial systems already exist for all of them:

- **Face-to-face window (*Face*)** systems are typically based on video-conference screens and cameras. Conventional videoconferencing software follows this paradigm, but in this article, we will refer to more immersive solutions involving natural-size appearance.
- **Remote Assistance (*Visit*)** applications allow a remote on-field person (the "host") to point a camera to the scene of interest so that the system user (the "visitor") can see it in real time[2]. Both users share the same point of view.
- **Shared Virtual Reality (*Meet*)** (also known as Shared Virtual Environment) are multi-user VR applications where each user is represented by an avatar in a common virtual space.

Three more archetypes cover two fundamental elements simultaneously. These systems are typically experimental, and are mostly described in the scientific literature, although some start-up companies also implement commercial prototypes:

- **Immersive Telepresence (*Face, Visit*)**. It is the extension of remote assistance applications by using an immersive camera (360) which is separate from the "host" user, so that the "host" is also seen in the video scene and therefore visual communication is possible (together with shared exploration of the remote environment).
- **Social Extended Reality (*Face, Meet*)**. It is the integration of real-time capture of a person, typically using volumetric video systems, into a shared virtual environment.
- **Immersive Digital Twin (*Meet, Visit*)**. It is an extension of a Shared VR experience where the virtual environment is a digital representation ("digital twin") of a physical location, at least from a visual representation perspective. Actions on the digital twin should have also effect on the physical world.

The **Distributed Reality** archetype covers all three elements simultaneously. So far, no system can cover the three of them in a significant way: Only some conceptual designs exist, as well as systems in fiction works, such as books or movies.

Archetypes which contain the *Visit* element are typically asymmetric: the "visitor" user feels immersed within the environment of the "host", but not the other way around. The reverse side of the communication ("visitor" to "host") normally follows one of the two paradigms:

- **AR Host.** The host user wears an AR HMD. Some visual cues are represented in the AR scene to represent the "visitor" user (i.e., there is a simple version of the "Meet" element), but not a full embodied avatar. Such cues can be also presented in regular phone or tablet displays or using projectors.

- **Audio Conference.** In those systems, all the communication is done through audio, although some visual information may also be included: list of participants, presence status for each of them, who is speaking, etc.

Finally, the **Move** element may be present (or absent) in any of the seven defined archetypes. However, the possibility to interact with the system is a key element of VR/MR/AR systems (Schäfer et al., 2021), and it enables *immersive collaboration* use cases (Apostolopoulos et al., 2012). Therefore, we consider it a fundamental element of immersive communication systems.

## 2.3 Components

Besides the different combination of fundamental elements they implement, immersive communication systems can be characterized by the technological components they are composed of (**Figure 3**):

- **Display**: Immersive systems typically use HMDs of any kind: AR, VR, or XR (VR display with attached camera to allow video pass-through). Large size screens, either 2D or 3D (light field displays) may also be used.
- **Avatar view** (how the user sees the other person) can be a Computer-generated imagery (CGI) model, animated by the application engine, or a real-time stream coming from a set of cameras. Different technologies are possible in either option.
- **World view** (how the user sees the surroundings of the other user). Again, the other user can be seen as part of an immersive video capture, or just immersed in a CGI environment, which may be a representation of the physical environment around the user or a completely virtual world. Some systems (e.g., AR) do not show any remote/virtual world at all.
- **Self-view** (how the user sees her/himself). VR/XR displays block the user direct self-view, resulting in the user not seeing him/herself, or getting a mediated view: Either an avatar representation or a video pass-through. AR HMDs and screens keep direct view of users' self.
- **HCI** interface (how the user can interact with the system). Different Human-Computer Interaction (HCI) technologies are possible: eye or head tracking, hand tracking/gestures, or different types of controllers.
- **Action.** Which *types* of actions are allowed in the system: whether they allow the user to move (locomotion), to interact with objects, or to point to specific locations. Besides, whether the action has *effect* in the physical world (e.g., by moving a physical robot), just in the virtual environment, or in a virtual representation of the physical world (twin).

Although any immersive communication system should, in principle, have most of these components, it is important to notice that *specific choices* of such components are normally used to implement the fundamental features elements before. *Face* and *Visit* elements are supported by real-time capture and transmission (i.e., "video") of the Avatar and Remote views, respectively. *Meet* element requires using an HMD Display. And *Move* is supported by Self-View (embodiment) and HCI interface (interaction). Possible action types and effects are also related to the fundamental elements implemented by the system, as it will be shown in **Section 2.5**.

---

[2]In the literature of remote assistance systems, the "visitor" is normally described as "remote" user, and the "host" is normally described as "local" user. However, since we are describing each side of the communication independently, we prefer a terminology which does not explicitly consider either side as being the "local" one.

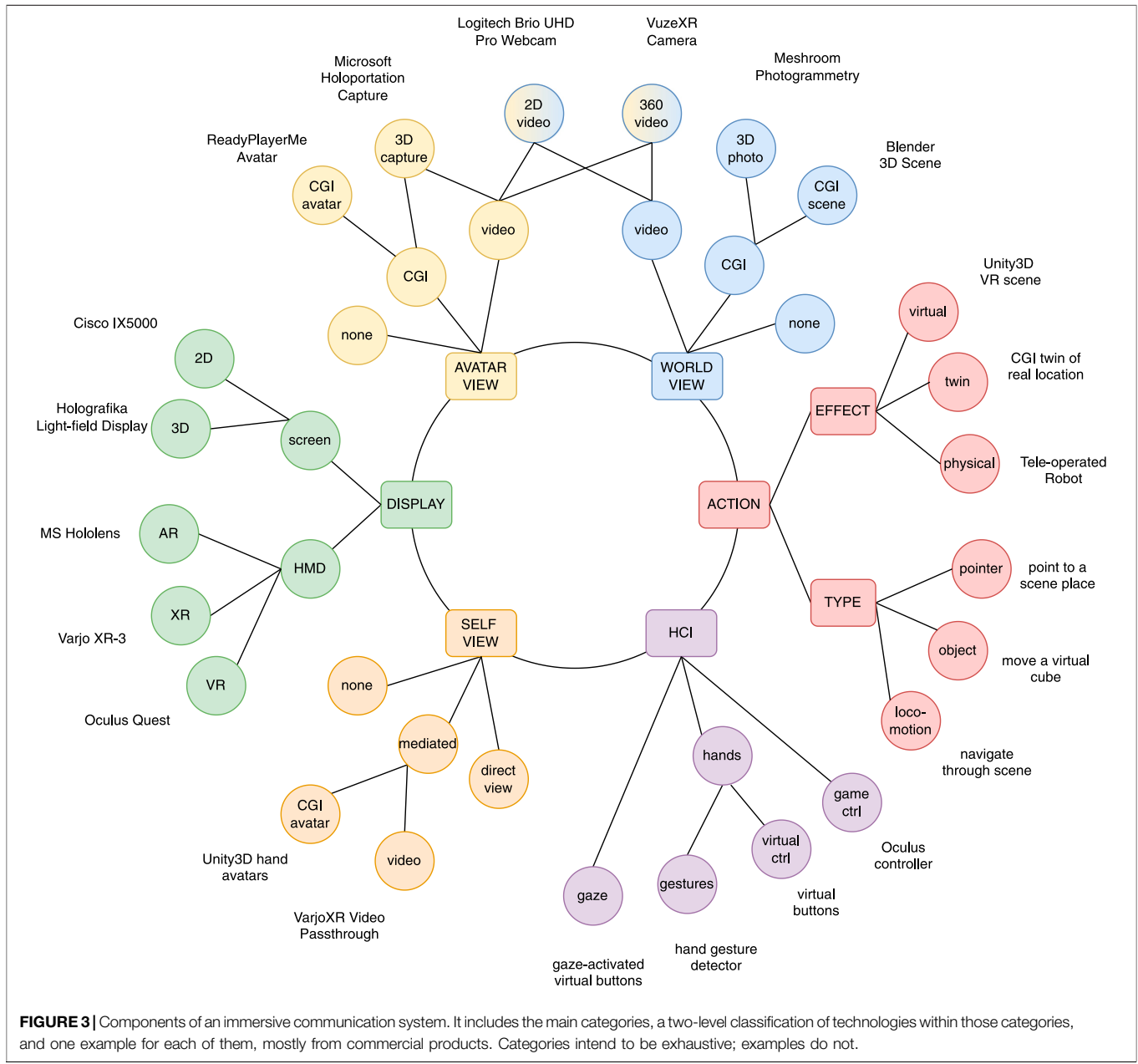

FIGURE 3 | Components of an immersive communication system. It includes the main categories, a two-level classification of technologies within those categories, and one example for each of them, mostly from commercial products. Categories intend to be exhaustive; examples do not.

## 2.4 Overview of Systems

This section intends to describe related works complying to the taxonomy presented in **Section 2**. It is our objective to focus exclusively on research or industrial attempts that go beyond the traditional videoconferencing systems by enhancing some of their limitations. However, this overview does not claim to be a complete state of the art review, but a selection of key representative systems that showcase the different archetypes of immersive communication systems. It also shows that most of the ICS existing in both the scientific literature and commercial services closely map to one of the seven system archetypes that we introduced, as depicted in **Figure 4** and that systems sharing the same archetypes usually involves the same subset of technological components, as depicted in **Figure 3**. As before, all system descriptions are done from the perspective of the receiver of the immersive communication.

### 2.4.1 Face-To-Face Window

Systems that prioritize **visual communications** that lie within **face-to-face window** archetypes mainly focus on achieving a very realistic representation of the user appearance and the counterpart dynamics conveying non-verbal communication cues. One of the first solutions of this kind was proposed in Kauff and Schreer (2002), namely 3D video-conferencing systems. This system was built around the idea of having meetings using a shared real/virtual table. At the real half of the table, local participants are sat down. The remote part of the table is captured and displayed using a 2D screen placed in such a way that it creates the illusion that the real table is extended with the remote table. Even though remote users are captured using conventional 2D video cameras, their efforts towards making the display look like an extended 3D space create the illusion of a 3D immersive system. Said 3D video conferencing system has later been converted into a product by Cisco, with their Cisco telepresence system Szigeti et al. (2009), that continues to be commercialized nowadays. More recently, Google Starline was proposed Lawrence et al. (2021), a more sophisticated and advanced system that enables 3D high resolution capturing using three synchronized stereo RGB-D cameras and audiovisual fidelity through the lenticular display that combines the different video streams using appropriate blending weights. This way, Google Starline ensures stereopsis, motion parallax, and spatialized audio. Other related factors and systems are under development within academic research groups, such as the visualization of the remote communication partner near the user gaze point (Kim S. et al., 2019), or real-time communication and exploration of the remote environment through Free Viewpoint Video systems (FVV) (Carballeira et al., 2022). All these systems have as their final aim to maximize the face-to-face communication. Here *screen displays* and *video avatars* are the key technological components involved, as they serve to represent the view of the user at the other side.

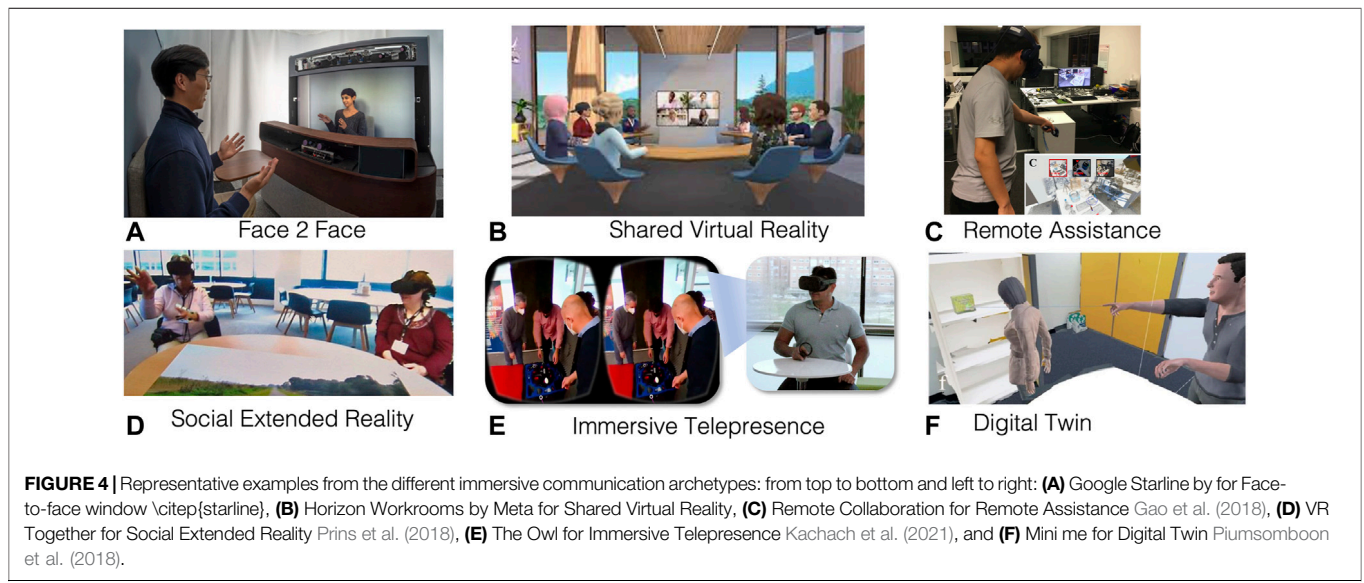

FIGURE 4 | Representative examples from the different immersive communication archetypes: from top to bottom and left to right: **(A)** Google Starline by for Face-to-face window \citep{starline}, **(B)** Horizon Workrooms by Meta for Shared Virtual Reality, **(C)** Remote Collaboration for Remote Assistance Gao et al. (2018), **(D)** VR Together for Social Extended Reality Prins et al. (2018), **(E)** The Owl for Immersive Telepresence Kachach et al. (2021), and **(F)** Mini me for Digital Twin Piumsomboon et al. (2018).

### 2.4.2 Shared Virtual Reality

**Shared Virtual Reality** is the archetype that focuses on **shared immersion** as fundamental element. The idea was already explored in the early days of Virtual Reality, as described in Durlach and Slater (2000), and today it is mature enough to be widely available commercially. This is the case of Shared VR platforms such as Mozilla Hubs from Mozilla, Virbela by Virbela, Facebook Spaces or Horizon Workrooms from Meta, AltSpace from Microsoft, to name a few. What all these platforms offer to users is the possibility to meet with other people in a virtual space, normally computer generated. Communication can revolve around a conversation on any topic, or it can revolve around the joint observation of an element (e.g., a video, a virtual object) within the scene, or the scene itself. User representations are normally computer generated but differ in the level of realism and personalization. Hubs use cartoonish avatars ranging from robots, animals, or human with combination of different demographics and face accessories that might be aligned with user's profile; Horizon Workrooms and Virbela allow to use as their virtual representation a fully customized avatar of their own person. Some academic works have also explored this type of systems, such as Pan and Steed (2017); Li et al. (2019). A detailed study on several Shared VR platforms can be found in McVeigh-Schultz et al. (2019). The principal technological components encompassing Shared Virtual Reality are: a *HMD display*, usually regular VR goggles, *Avatar*, in the form of CGI or cartoonish. Self-view might be used, but always in the form of CGI avatar representing the entire body, or just the hands. The *world* that users shared is usually represented as a CGI scene. *HCI* interfaces such as eye, head and/or hand tracking are also allowed for interaction with the virtual world. *Actions* related to movement and effects in virtual scenes are also allowed.

### 2.4.3 Social Extended Reality

**Social Extended Reality** is a system archetype that addresses both **visual communication** and **shared immersion**. Microsoft Holoportation system developed by Orts-Escolano et al. (2016) and its foreseen evolution Mesh (Microsoft, 2021), as well as the VR Together solution studied in Gunkel et al. (2018); Li et al.

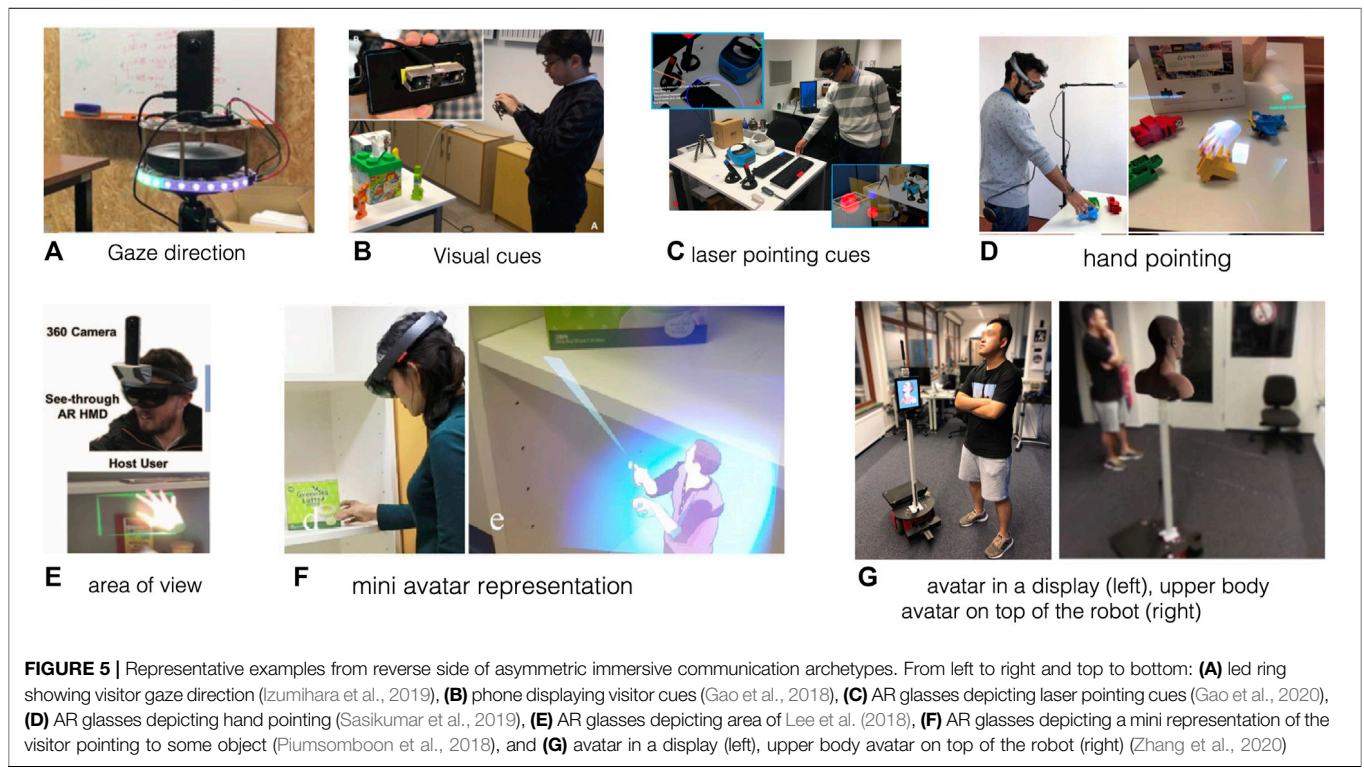

**FIGURE 5 |** Representative examples from reverse side of asymmetric immersive communication archetypes. From left to right and top to bottom: **(A)** led ring showing visitor gaze direction (Izumihara et al., 2019), **(B)** phone displaying visitor cues (Gao et al., 2018), **(C)** AR glasses depicting laser pointing cues (Gao et al., 2020), **(D)** AR glasses depicting hand pointing (Sasikumar et al., 2019), **(E)** AR glasses depicting area of Lee et al. (2018), **(F)** AR glasses depicting a mini representation of the visitor pointing to some object (Piumsomboon et al., 2018), and **(G)** avatar in a display (left), upper body avatar on top of the robot (right) (Zhang et al., 2020).

(2021), are examples of these immersive communication systems that maximize both user realism and the possibility to see things together virtual. Differing from the given examples of pure visual communication, here the use of a Virtual Reality or Augmented Reality HMDs are required. Also, unlike the given examples of pure shared immersion, the realism of users is important. Earlier approaches simply inserted a 2D video of the remote user within the AR or VR environment, as in Lawrence et al. (2018); Prins et al. (2018). In more advanced implementations, such as the aforementioned Holoportation or VR Together, users are represented using 3D point clouds obtained from volumetric video captures. In case there are spatial constrains for full-size avatar placement, the use of miniature versions is becoming a common practice (Piumsomboon et al., 2018), yet eye contact needs to be carefully designed (Anjos et al., 2019). The components involved in Social XR are similar to those present in Shared XR, but with some subtle differences: *HMD displays* tend to be more AR or XR goggles, *avatars* are based on video or volumetric capture to increase their realism; *worlds* might either be CGI scene or scenes based on photogrammetry, or none, as in the case of Holoportation. All type of *actions* related to object manipulation, pointing and locomotion are supported and referred to the virtual world

### 2.4.4 Remote Assistance

Immersive communications focusing on exploration tasks benefit more from the use of **remote assistance** systems. In this context, the user (the "visitor") provides practical expertise to the remote use (the "host") who is physically located in an area of interest. Usually a 360° video camera is capturing the area of interest that is transmitted so that the user wearing HMD can explore it. Jack in Head was proposed in Kasahara et al. (2017) and represents a telepresence system in which the host is wearing a head-mounted omnidirectional camera, composed of nine cameras around the user's head and backpack containing a laptop computer. As already mentioned, system archetypes which contain the Visit element are typically asymmetric (**Figure 5**). In the particular case of Jack in Head, the reverse side, communication that the host received from the visitor, is purely based on audio cues.

The work by Wang et al. (2020) relies on the use of a 2D conventional camera capturing at 1280, ×, 960 pixels. Although not omnidirectional, the 2D camera feed is rendered in Unity 3D in a particular plane, allowing the user to see it while wearing the HMD. In this case, the reverse side falls into the category of the AR Host archetype yet using a standard projector and not AR goggles. In this work, apart from audio communication, the host might have further visual cues such as cursor pointer, head pointer or eye gaze from the visitor.

Most related works dealing with remote assistance rely on the use of a commercial 360° video cameras placed on the head of the host. Representative examples are the works in Lee et al. (2018); Teo et al. (2019a); Young et al. (2019), who proposed the use of a Ricoh Theta S 360° to capture a 360° video and stream it to the visitor. For those use cases where the area of interest is mainly static, there are also hybrid immersive approaches resulting from the combination of 3D reconstructed environment and live 360° video panorama camera (Teo et al., 2019b, 2020). In turn, all these systems share the use of the AR Host archetype with AR googles as the mainstream system for the reverse side. In particular, AR google can render visual cues made by the user such as: gaze, user awareness, hand gestures (Teo et al., 2019a) (easily inferred through the use of hand trackers such as Leap Motion), hand pointing (Kim et al., 2020) or annotations through the use of VR controllers (Teo et al., 2019a).

The key components involved in Remote Assistance is the use of VR HMD as *displays*; the use of 360° camera to capture the world environment. As usually the camera is placed on the host' head, it is not possible for the visitor to see the any visual representation of the host. Also, the visitor might have gaze, or pointer towards the physical space of the host. At the reverse side of the communication the host is not using any of the key technological components depicted in **Figure 3**, but simply audio cues.

One important concern related to these systems is how to reduce the cybersickness resulting from the movement of the host wearing the omnidirectional camera. As reported by several authors such as Singla et al. (2017); Pérez et al. (2018), the onset of cybersickness in 360 video is mostly triggered by the accelerated motion of the capture camera. In this sense, the system by Kasahara et al. (2017) proposed the use of some image-processing algorithm to diminish the rotational motion and achieve therefore real-time camera stabilization and cybersickness alleviation.

### 2.4.5 Immersive Telepresence

In the intersection between **remote assistance** and **face-to-face window** archetypes lie **immersive telepresence** systems. This type of immersive systems are characterized by the use of a 360° camera that, in contrast to remote assistance systems, is not placed on the user's head (which allows having the same vision as the user), but rather uses a tripod, held by the hands of a user, or even placed on top of a robot (Zhang et al., 2020). While communication happening in **remote assistance** or **face-to-face window** systems is normally centered between two communication partners, immersive telepresence systems are designed for multi-user use cases, with a minimum of two and up to N communication partners, normally limited by the technological limitations of the system. As an evolution of remote assistance systems, *Giant miniature collaboration*, the system prototype proposed by Piumsomboon et al. (2019), aims to connect visitor and host, while the host carries the 360° camera on his/her hand, which allows the user to see the face of the host.

*Show me around*, the system proposed by Nassani et al. (2021), has been designed for immersive Virtual Tours. One of the participants in the area of interest holds a Ricoh Theta V 360° camera connected laptop computer with audio headphones and a microphone to capture the whole scene where the guide is also present. This 360° video is integrated into the open source video conferencing platform Jitsi, which allows a large number of users (20, – ,30) to access this immersive content through a 2D screen, although without using VR goggles. Likewise remote assistance system, immersive telepresence systems are asymmetric. In the work by Nassani et al. (2021), the reverse side is within the classification of AR host, as the remote user physically co-located with the 360° camera, can see through a screen, 2D videos of every user connected to the session (as in regular 2D videoconferencing), along with hand pointing cues that are overlapped to the also displayed 360° scene.

*Exleap*, an immersive telepresence system done by Izumihara et al. (2019) gives users the possibility to move around the area of interest through the use of different 360° cameras, which are placed at strategic positions, creating the experience of leaping between places. To achieve this, only VR goggles are required. At the reverse side, the communication is purely based on audio along with a led string wrapping the camera node to indicate existence of users and respective direction. *Owl* is also an immersive telepresence system developed by Kachach et al. (2021). It is a low cost prototype consisting of either Ricoh Theta V 360° or Vuze XR as omnidirectional camera along with Raspberry Pi with a touchscreen, a standard hands-free speaker, a 4G/5G modem, and a power bank, all placed in a custom 3D-printed housing. With this system different users joining the session can experience the real space as if they were there using VR goggles such as Quest2. Alternatively, Vuze XR can also work in 180° stereo mode, providing the user depth perception. Further, they have the possibility to see avatars of other users joining the session, as well as content such as PDFs or images while being immersed. However, this system archetype is purely based on audio communication. In this regard, there are other related works such as *Think Fast* (Zhang et al., 2020) that represent the visitor depicting his avatar on a tablet attached to the 360 camera, or even by rendering his/her upper body on top of the robot so that the host can see him/her using AR glasses.

Recent works also offer the possibility of 3D panorama streaming. Bai et al. (2020) developed a 3D panorama sensor cluster that supports instant 3D reconstruction with real-time updating. This is done through a cluster of assembled eight off-the-shelf RGB-Depth cameras into a sensor cluster, which are correctly aligned to create a semi-sphere with no gaps. At the reverse side, an AR host system is found. Gaze is depicted as a virtual ray cast line and hand gestures are represented using a 3D mesh of the user's hand.

The components involved in Immersive Telepresence slightly changes from those from Remote Assistance archetype. *Displays* might be XR HMD. As the camera is not placed on the host' head, there is the possibility for the visitor to see the host through the *world representation* based on 360° video. Besides, in what concern *actions*, the user can point to the real place of the host, and experience (partial) locomotion through for instance, switching between cameras, or being able to move by placing the camera into a robot.

### 2.4.6 Immersive Digital Twin

**Digital Twin** archetypes make use of visit and meet elements. This is one of the most challenging systems as it implies that two users are feeling at the same time visiting a real place while having the possibilities of seeing each other through avatar representations. The closest related work a is the asymmetric system by Piumsomboon et al. (2018), where a typical remote assistance system is enhanced by the possibility of feeling immersed in a real space with another user on it being represented by a 3D Mesh.

TABLE 2 | Implementation of the immersive communication system archetypes.

| Archetype | Elements[a] | Display | Avatar View | World View | Action[b] |
| --- | --- | --- | --- | --- | --- |
| Face-to-face Window | F | 2D/3D Screen | 2D/3D Video | N/A | — |
| Shared VR | M | VR HMD | CGI | CGI | O, P, L (virt) |
| Remote Asssistance | V | VR HMD | N/A | 2D/360 Video | P (twin/phy) |
| Social XR | F, M | HMD | 2D/3D Video | CGI | O, P, L (virt) |
| Immersive Digital Twin | M, V | VR HMD | CGI | 3D Photo | O, P (twin) |
| Immersive Telepresence | V, F | HMD | 360 video | | P, L (phy) |
| Distributed Reality | F, M, V | Concept Only | | | |
| AR Host | M (simple) | AR HMD | CGI cues | N/A | P (twin) |
| Audio conference | — | Screen | Icons | — | — |

[a]F(ace), V(isit), M(eet).
[b]O(bject), P(ointer), L(ocomotion); phy(sical), virt(ual), twin.

Digital Twins involves many different technological components: a VR or XR goggle as *HMD display*; a *world representation* that attempts to create a digitized (or twin) version of a real place. The receiver of the communication sees a representation of the other user as part of this *world representation*. This user also has the possibility to point and manipulated objects in the twin space. At the reverse side of the communication, the user at the real local place use *display*, in the form of AR goggles that enables to see a volumetric representation of the user at the other side.

### 2.4.7 Distributed Reality

Ideal immersive communication systems should be able to simultaneously provide shared immersion, visual communication, and remote presence: The ability for one user to virtually "teleport" to the location of the other, feel immersed in the same location and interact with him/her face to face. This paradigm is defined as **Distributed Reality** by Villegas et al. (2019). The concept is widely depicted in fiction movies, such as the Jedi Council of Star Wars prequel trilogy, but no practical implementation exists already.

## 2.5 A System: Its Fundamental Elements, Archetype, and Components

Based on **Section 2.4**, all reported ICS systems can be classified into one of the seven proposed system archetypes. Besides, after analyzing all the systems per archetype, we observed that those pertaining to the same archetype tend to share the same key technological components. As a summary, **Table 2** defines each archetype based on the key fundamental elements involved as well as their implementation components.

Self-view and HCI components, mostly related to the fundamental element *Move* have not been included in the table. There is a basic dependency between the display technology and self-view: screen and AR HMD displays allow direct self-view of the user, while fully occlusive HMDs (VR/XR) impose a mediated representation, or no self-view at all. Except of that (relevant) restriction, different self-perception and interaction technologies can be integrated with any system, independently of the communication paradigm which is applied.

Likewise with existing ICS, better insights will be driven if upcoming ICSs are also characterized in terms of their fundamental elements, system archetype and key technological components.

## 3 EXISTING QOE METHODOLOGIES FOR IMMERSIVE COMMUNICATIONS

In this section, we aim to review from the QoE perspective the related works presented in **Section 2.4** from the system point of view. Firstly, we present some background concepts of QoE and, especially, on presence, which is the most relevant feature addressed in ICS studies. Then, the following subsections provide an overview of the considered related works, which share a common basic design: Two or more users communicate through the system and perform one or several **communication tasks**. The same task is performed under two or more different **experimental conditions**, which are normally related to using different communication systems (e.g., comparing an immersive with a non-immersive one) or different configurations of the same system. Finally, several **QoE features** are evaluated, studying whether there exist significant differences between the experimental conditions. Therefore, for the sake of clarity, the overview of the related works is presented in terms of the covered features, tasks, and conditions.

The analyzed related works are summarized in **Table 3**. Additionally, there are some few works which are focused more on interaction rather than on communication (Brunnström et al., 2020; Villegas et al., 2020). Still, it is still interesting to understand their related evaluation methodologies for hybrid immersive systems where both communication and interaction are important. For more details regarding the questionnaires reported in **Table 3**, please refer to **Table 4**.

### 3.1 Background

Immersive technologies are introducing new factors that influence the QoE of the end users in comparison with previous technologies (Perkis et al., 2020). For instance, following the common classification of influence factors (Le Callet et al., 2012), the QoE of the users of ICS can be

**TABLE 3** | Related works on QoE evaluation for the different archetype systems. CT stands for Completion Time; AE stands for Angular Error; HM for Head Movements, HG for Hand Gestures, HN for Head Nodes, EM for eye movements.

| Work | Archetype | Context | Dialogue | Exploration | Manipulation | Conditions | High level feature | Low level Feature |
|---|---|---|---|---|---|---|---|---|
| Kim et al. (2019b) | Face2Face window | Watch a movie | Comment on the movie | — | — | Big display/small display with following gaze/corner display | Subscale from NMM-SPI for emotions Likes/dislikes about gaze following | — |
| Lawrence et al. (2021) | Face2Face window | Conferencing | Semi structured conversation with a research confederate | — | — | 3D video conferencing vs. 2D video conferencing system | HOLO for presence, attentiveness, connectedness, reaction-gauging | HG, HN, and EM for non-verbal behaviour |
| Pan and Steed, (2017) | Shared VR | Play games | — | — | Solving puzzles (pieces) | Embodiment types: no self-avatar, self-avatar, and face2face | IT for trust | CT for task performance |
| Li et al. (2019) | Shared VR | Photo sharing | Comment on shared photos | — | — | Face2face, skype, and Facebook Spaces | SocialVR for social presence PMRI for emotions | — |
| Orts-Escolano et al. (2016) | Social XR | Play and collaborate remotely | Tell a lie game and dialogue to build blocks | — | Building blocks | VR vs. AR | Semi-structured interview for: presence, interaction, exploration, etc. | Semi-structured interview for: visual quality and latency |
| Gunkel et al. (2018) | Social XR | Watch a movie and play a game | Comment on the movie or game | — | — | System performance | RGQ: Social presence, interaction, exploration, and global QoE | RGQ: visual and audio quality |
| Prins et al. (2018) | Social XR | Play a game | — | — | Pong (two-player game) | System performance | Feedbak on presence and overall quality | — |
| Lawrence et al. (2018) | Social XR | Conferencing | Negotiation | — | Assembly of lego blocks | Audio only, video fixed to HMD, and video fixed to host world | NMM-SPI | — |
| Li et al. (2021) | Social XR | Watch a VR Movie (inside the scene) | Questions raised by movie characters | Follow Characters | Interact with environment (e.g., click buttons) | HMD vs. screen with game controller | WS for presence SocialVR for social presence SSQ for cybersickness NASA-TLX for mental workload | VQoE for visual quality |
| Lee et al. (2018) | Remote assistance | Remote collaboration | — | Find a set of target objects in the task space | Place the target objects on the desk | Dependent view vs. independent view | MEC spatial for presence NMM-SPI for social presence SMEQ for workload SSQ for cybersickness | CT for task performance |
| Young et al. (2019) | Remote assistance | Remote exploration | — | Explore remote environment | – | Three ways of interaction | IPQ for presence NMM-SPI for social presence and emotions SSQ for cybersickness | — |
| Teo et al. (2019a) | Remote assistance | Remote collaboration | — | Identify objects | Decorate a bookshelf placing objects | No cues, hand gestures, pointer, and hand gestures + pointer | MEC for spatial presence NMM-SPI for social presence SMEQ for workload | SUS for usabilty |

TABLE 3 | (*Continued*) Related works on QoE evaluation for the different archetype systems. CT stands for Completion Time; AE stands for Angular Error; HM for Head Movements, HG for Hand Gestures, HN for Head Nods, EM for eye movements.

| Work | Archetype | Context | Dialogue | Exploration | Manipulation | Conditions | High level feature | Low level Feature |
|---|---|---|---|---|---|---|---|---|
| Teo et al. (2020) | Remote assistance | Remote collaboration | — | Find a set of target objects in the task space | Place the target objects to a specific location | 360 image mode vs. 360 projection mode | MEC for spatial presence NMM-SPI for social presence SSQ for cybersickness SMEQ for workload SEQ +3 custom questions for global QoE | SUS for usabilty CT for task performance |
| Wang et al. (2020) | Remote assistance | Remote collaboration on physical tasks | — | Locate blocks and follow remote pointers | Assembly of lego blocks | Cursor pointer, head pointer, and eye-gaze pointer | TQ: co-presence, interactivity, and exploration NASA-TLX for mental workload | TQ: Visual and audio quality |
| Bai et al. (2020) | Remote assistance | Work together remotely | — | Search blocks and follow remote indications and visual cues | Assembly of lego blocks | Verbal only, eye gaze, hand gesture hand gaze + hand gesture | MEC for spatial presence NMM-SPI for social presence NASA-TLX for mental workload | SUS for usabilty CT for task performance |
| Anjos et al. (2019) | AR host | Play games. | Solve riddles | — | — | System performance | — | Semi-structured interview for task performance |
| Kasahara et al. (2017) | Immersive telepresence | Cleaning up a lab room | — | Locate objects and clean them | — | Video Stabilization | SSQ for cybersickness | HM for non-verbal behaviors |
| Piumsomboon et al. (2019) | Immersive telepresence | Remote collaboration | Guess objects of interest | House inspection | Arrange objects | Types of virtual representations, levels of miniature control, levels of 360-video view dependencies, and 360-camera placement positions | MEC for spatial presence NMM-SPI for social presence SSQ for cybersickness SMEQ for workload | SEQ for task performance |
| Zhang et al. (2020) | Immersive telepresence | Telepresence | — | Locate/indicate remote user's gaze | — | Distance to avatar and display (AR, tablet) | — | AE for task performance |
| Piumsomboon et al. (2018) | Immersive digital twin | Remote collaboration | — | Identify objects | Place objects to a specific location | Fixed life-size full-body avatar with and without Mini-me | NMM-SPI for social presence SMEQ for workload | SEQ for task performance |
| Brunnström et al. (2020) | No communication | Remote control | — | — | Control a crane to load logs | Latency | RC: Comfort, immersive, and overall quality | RC: Picture, responsiveness, and task accomplishment quality |
| Villegas et al. (2020) | No communication | Escape room game | — | — | Manipulate game objects | Real hands vs. VR controllers | WS for presence Embodiment DREQ for Global QoE | — |

influenced by new human factors, such as the possibility to freely explore the immersive content, system factors, such as the use of HMDs, and context factors, such as the sharing social experiences in immersive environments.

The influence factors can be described by users in terms of perceptual features (Le Callet et al., 2012), which are related to the characteristics of the individual experience and can be classified into different levels. For instance, the Qualinet white paper on definitions of QoE, defined four levels: direct perception, interaction, usage situation, and service. Another example is provided by Chen et al. (2012) (and standardized in the ITU-R recommendation BT. 2021 (ITU, 2015)) for stereoscopic 3D content, in which primary (picture quality, depth quality, and visual discomfort) and secondary (naturalness and sense of presence) features were defined.

TABLE 4 | Description of common questionnaires used in immersive communication experiments.

| Questionnaire | Measure, N. items, Scale, Subscales, Factors |
|---|---|
| Interpersonal Trust (IT) Johnson-George and Swap (1982) | Interpersonal trust in social situations<br>21 items for male version, 13 items for female version<br>9-point Likert scale.<br>Subscales: reliableness (male and women), emotional trust (male and women), and general trust (male) |
| Nasa Task Load Index (NASA-TLX) Hart and Staveland, (1988) | Workload of task<br>6 items<br>21-point Likert scale<br>Measuring mental, physical and temporal demand, performance, effort and frustration. |
| Simulator Sickness Questionnaire (SSQ) Kennedy et al. (1993) | Users' levels of cybersickness symptoms<br>16 items<br>4-point scale<br>3 subscales: Nausea (N), Oculomotor (O), and Disorientation (D) |
| Subjective Mental Effort Question (SMEQ) Zijlstra, (1993) | One question that measures the mental effort<br>1 item<br>9 labels scale (from "Not at all hard to do" to "Tremendously hard to do") |
| Witmer and Singer (WS) Witmer and Singer, (1998) | Sense of presence<br>32 items<br>7-point Likert scale.<br>Subscales: Involvement/Control, Natural, Auditory, Haptic, Resolution, Interface Quality<br>Factors: Involvement/Control, Natural, Auditory, Haptic, Resolution |
| System Usability Scale (SUS) Brooke, (1996) | Usability<br>10 items<br>5-point Likert scale. |
| Igroup presence questionnaire (IPQ) Schubert et al. (2001) | Presence<br>14 items<br>–4<br>Subscales: spatial presence, involvement, experienced realism |
| Networked Minds Measure Social Presence Inventory (NMM-SPI) Harms and Biocca, (2004) | Social presence and emotions<br>36 items<br>9-point Likert scale<br>Subscales: co-presence, attention allocation, perceived message understanding, perceived affective understanding, perceived emotional independence, and perceived behavioral independence |
| MEC Spatial Presence Questionnaire (MEC-SPQ) Vorderer et al. (2004) | Spatial presence<br>32, 48, 64, items (4, 6 and 8 items per each of the 8 subscales), 5-point Likert scale<br>Subscales: attention allocation, higher cognitive involvement, suspension of disbelief, spatial situation model, spatial presence self location, spatial presence possible actions, domain specific interest, visual spatial imagery |
| Single Ease Question (SEQ) Sauro and Dumas, (2009) | Assessment of how difficult users find a task<br>1 item<br>7-point Likert scale |
| Pictorial Mood Reporting Instrument (PMRI) Vastenburg et al. (2011) | Emotions<br>9 items<br>5-point scale<br>Nine moods: excited, cheerful, relaxed, calm, bored, sad, irritated, tense and neutral |
| Embodiment Gonzalez-Franco and Peck, (2018) | User embodiment on immersive experiences<br>25 items<br>7-point Likert scale<br>Subscales: body ownership, tactile sensations, location of the body, external appearance, and response to external stimuli |
| Feedback Prins et al. (2018) | Requirements gathering to understand user expectations for social VR.<br>6 items<br>7-point Likert scale<br>Subscales: Social presence, interaction, exploration, visual quality, audio quality, and overall quality |
| Requirements Gathering Questionnaire (RGQ) Gunkel et al. (2018) | Short questionnaire to gather feedback form users' immersive experiences<br>2 items<br>5-point Likert scale<br>Subscales: Presence and overall quality |
| SocialVR Questionnaire Li et al. (2019) | Social and interactive experiences in immersive media<br>24 items<br>5-point Likert scale<br>3 subscales: Presence/Immersion (PI), Social Meaning (SM), and Quality of Interaction (QoI) |

TABLE 4 | (*Continued*) Description of common questionnaires used in immersive communication experiments.

| Questionnaire | Measure, N. items, Scale, Subscales, Factors |
|---|---|
| Distributed Reality Experience Questionnaire (DREQ) Perez et al. (2019) | Presence and quality aspects<br>10 items<br>5-point scale<br>Subscales: presence, video quality, cybersickness and quality of experience |
| Remote Control (RC) Brunnström et al. (2020) | QoE aspects<br>6 items<br>5-point Likert scale<br>Subscales: picture quality, comfort quality, immersive quality, overall quality, responsiveness quality, and task accomplishment quality |
| Tele-collaboration Quality (TQ) Wang et al. (2020) | Social presence (slightly modified from Gupta et al. (2016) and Harms and Biocca, (2004) to better reflect the experiment)<br>7 items<br>7-point Likert scale<br>Subscales: co-presence, interactivity, exploration, visual quality, audio quality, fatigue |
| Visual Quality of Experience (VQoE) Li et al. (2021) | Visual quality of self and others' volumetric representations<br>2 items<br>5-point Likert scale |
| Holoportation Questionnaire (HOLO) Lawrence et al. (2021) | Global QoE for holoportation systems<br>7 items<br>5-point scale for 6 items and 7-point scale for one<br>Subscales: presence, attentiveness, personal connection, reaction-gauging, engagement, closeness, eye-contact |

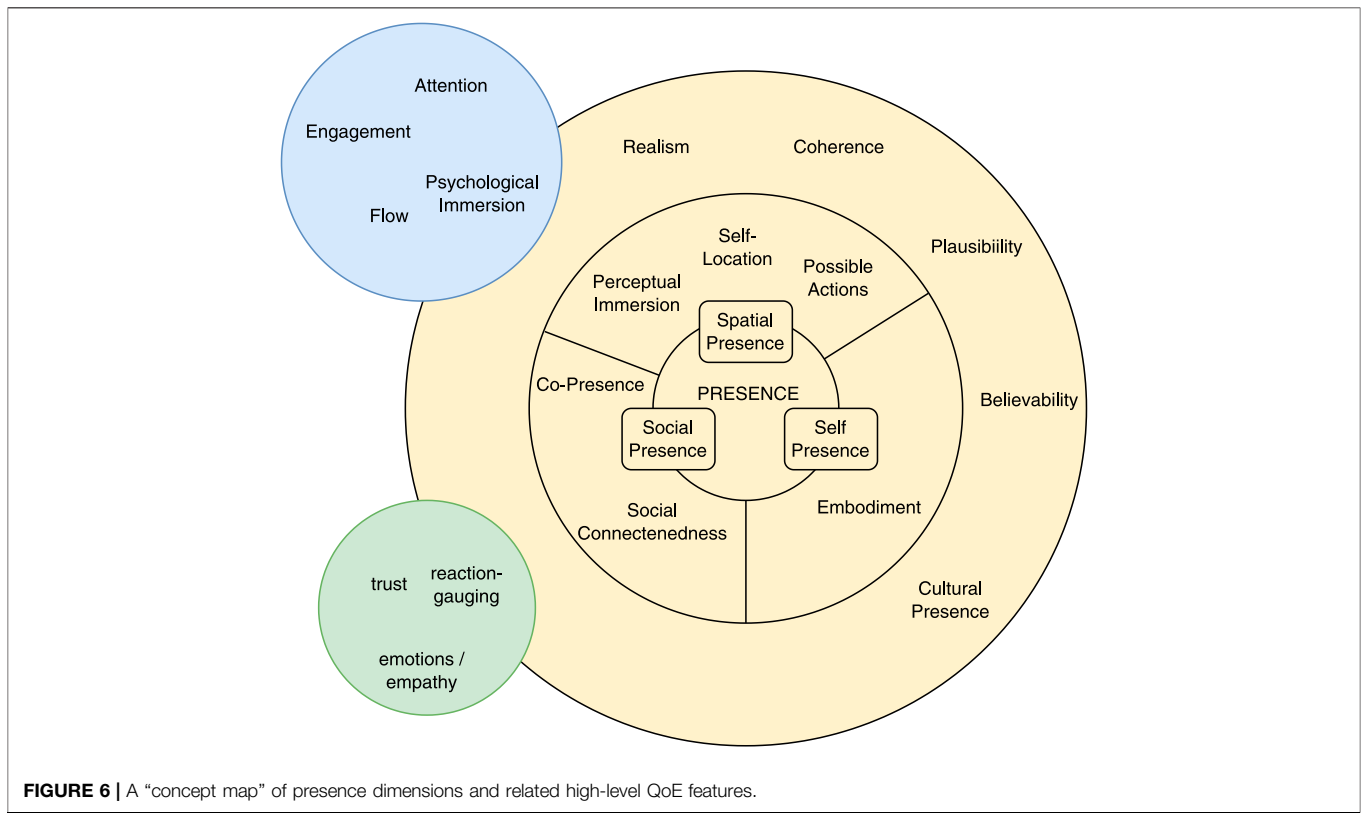

FIGURE 6 | A "concept map" of presence dimensions and related high-level QoE features.

With the emergence of immersive technologies and in order to cover the new features involved in the immersive experiences, the QoE research is embracing concepts and methods from the human-computer interaction and user experience communities, especially in terms of the evaluation of user-centric features related to the immersion and sense of presence. Thus, we have identified **high-level** features, which focus on the **user**, addressing cognitive or psychological constructs such as presence, empathy or workload; and **low-level** features, which directly evaluate technical aspects of the **system**: audio-visual quality, responsiveness, usability, etc.

From the overview of the related works detailed in **Section 3.2**, different high-level features have been considered when evaluating the QoE of the users of ICS. Among them, the most relevant features are related to presence, defined as "the subjective experience of being in one place or environment, even when one is physically situated in another" (Witmer and Singer, 1998). The phenomenon of presence is complex and multi-factorial, as shown in the graphic representation in **Figure 6**. For a thorough discussion on presence, the reader can refer to Skarbez et al. (2017) or Lombard and Jones (2015), among others.

Our starting point is the analysis in Biocca (1997), which identifies three forms in which users can feel present in a virtual environment: **Spatial Presence** (the sense of "being physically there"), **Social Presence** ("being with another body"), and **Self Presence** ("this body is really me").

**Spatial Presence** is frequently analyzed in two dimensions (Hartmann et al., 2015):

- *Self-location*: The feelings that the user has departed from their real environment and feel they are in other place. Also defined as Place illusion by Slater (2009). It is mostly achieved through "perceptual immersion" (Lombard et al., 2000): The ability of the XR system to immerse the user into a different environment.
- *Possible actions*: Users' subjective impression that they would be able to carry out actions in the environment. A similar view is the Plausibility Illusion (Slater, 2009), the illusion that the depicted scenario is actually occurring.

**Social Presence** has two dimensions or perspectives (Lee, 2004; Skarbez et al., 2017):

- *Co-presence*, or the sense of being together with another or others, the "condition in which instant two-way human interactions can take place" (Zhao, 2003). It is a perceptual feature: co-presence exists whenever a person senses that there is another person in the environment.
- *Social connectedness*, "social presence illusion" or simply "social presence", the engagement with others (which requires interaction), defined by Van der Land et al. (2011) as "the awareness of being present with others in a mediated environment combined with a certain degree of attention to the other's intentional, cognitive, or affective states".

**Self-presence** implies the self-perception of one's body, emotions and/or identity (Ratan, 2013). The perception of one's body, or embodiment, has some dimensions on its own, as described by Kilteni et al. (2012): *self-location* (being inside a body), *agency* (having global motor control of the body), and *body ownership* (self-attribution of a body).

There are other aspects of presence which are not totally included in our previous classification, and somehow overlap among them (Lombard and Jones, 2015; Skarbez et al., 2017):

- *Realism, plausibility, coherence, cultural presence.* Those concepts describe different ways in which the experience feels "real" or "realistic", shows internal coherence, matches with the users' expectations and cultural background, etc. Most authors agree that these factors form an integral part of the concept of presence.
- *Psychological immersion, engagement, attention, flow.* Those concepts describe the situation when the user feels absorbed, engaged or involved with the environment. Although they are closely related to presence, most authors agree that these factors describe a different psychological construct.
- *Empathy, emotion conveyance, trust, etc.* Although a relation with social presence exists, those concepts describe a different phenomenon, more related to the area of empathic communication.

## 3.2 Features

Among the features evaluated in the user studies of the reviewed works, the most common one is presence, evaluated in terms of spatial, social, and self-presence. As shown in **Table 3**, presence was considered in user studies involving almost all system archetypes, such as Social XR (Orts-Escolano et al., 2016; Gunkel et al., 2018; Lawrence et al., 2018; Prins et al., 2018; Li et al., 2021), Shared VR (Li et al., 2019), face-to-face window (Lawrence et al., 2021), remote assistance (Lee et al., 2018; Teo et al., 2019a, 2020; Young et al., 2019; Bai et al., 2020; Wang et al., 2020), immersive telepresence (Piumsomboon et al., 2019), and digital twin systems (Piumsomboon et al., 2018). As reflected in **Table 4**, different questionnaires have been used to evaluate presence in its different forms, although some traditional ones are used in several different studies, such as the one proposed by Witmer and Singer (WS) (Witmer and Singer, 1998) and the MEC Spatial Presence Questionnaire (MEC-SPQ) from Vorderer et al. (2004) for spatial presence, and the Networked Minds Measure Social Presence Inventory (NMM-SPI) from Harms and Biocca (2004) for social and co-presence. While some other studies used variations of these questionnaires [e.g., (Young et al., 2019)] used modified versions of NMM-SPI for co-presence and of the Igroup presence questionnaire (IPQ) from Schubert et al. (2001) for spatial presence, other works proposed their own questionnaires. For instance, to assess spatial presence Li et al. (2019) proposed the SocialVR questionnaire, Lawrence et al. (2021) the Holoportation questionnaire (HOLO), and Wang et al. (2020) the Telle-collaboration Quality (TQ), Gunkel et al. (2018) the VR experience questionnaire (with a 7-point scale question among four questions related to the VR experience), and Prins et al. (2018) (with a single 5-point scale question) for social presence.

After presence, workload and simulator/cyber sickness are factors commonly evaluated in the considered systems, mainly using well-established questionnaires, such as the Nasa Task Load Index (NASA-TLX) (Hart and Staveland, 1988) and Subjective Mental Effort Question (SMEQ) (Zijlstra, 1993) questionnaires for workload, and the Simulator Sickness Questionnaire (SSQ) (Kennedy et al., 1993). As mental stress conditions have an effect in the Central Nervous System (CNS) or the Autonomic Nervous System (ANS), it is also possible to monitor them using psychophysiological measures (Engelke et al., 2016), even though none of the works reported in **Table 3** uses this approach. In other contexts, for instance, Electroencephalography (EEG) has been used to detect mental load (Jiao et al., 2018) or cybersickness (Kim J. et al., 2019) in VR users.

The SocialVR questionnaire proposed by Li et al. (2019) included ten items to evaluate quality of interaction in Li et al. (2021) and eleven items covering social connectedness. Similarly, these two factors were also evaluated by Wang et al. (2020) using two of the subscales included in their TQ questionnaire. In addition, Gunkel et al. (2018) included one item in their VR experience questionnaire to evaluate interactivity and Lawrence et al. (2021) evaluated social connectedness with one subscale of their HOLO questionnaire.

Other factors were evaluated in a few studies. For example, Kim S. et al. (2019) used questions adopted from the NMM-SPI questionnaire Harms and Biocca (2004) to evaluate sharing emotions with their face-to-face window system, while Li et al. (2019) used a self-report emotion rating questionnaire proposed by Vastenburg et al. (2011). Also, exploration and the global QoE was evaluated with two items of the VR experience questionnaire of Gunkel et al. (2018), attentiveness and reaction/gauging were evaluated by Lawrence et al. (2021) with two subscales of their HOLO questionnaire, Pan and Steed (2017) evaluated trust using the questionnaire proposed by Johnson-George and Swap (1982), Brunnström et al. (2020) analyzed immersion and comfort (among other factors) with their RC questionnaire, and Villegas et al. (2020) studied embodiment with the questionnaire proposed by Gonzalez-Franco and Peck (2018).

Regarding lower level features more related to technical factors of the immersive communication systems, task performance is the most commonly evaluated in user studies, both through objective measurements and questionnaires. For example, completion time seem to be one of the most relevant measurements (Pan and Steed, 2017; Lee et al., 2018; Bai et al., 2020; Teo et al., 2020) to assess task performance, while other technical measurements more specific to the task under study can also be used [e.g., angular error in the estimation of the location of the remote user's gaze in Zhang et al. (2020)]. An example of useful questionnaires to evaluate task performance is the Single Ease Question (SEQ) (Sauro and Dumas, 2009) used by Piumsomboon et al. (2018, 2019); Teo et al. (2020). Finally, semi-structured interviews can also be considered, as well as preference rankings to compare the performance among the considered conditions in Kim S. et al. (2019). Visual quality is another factor commonly evaluated in related user studies, mainly using questionnaires. For example, Li et al. (2021) proposed the Visual QoE (VQoE) questionnaire to rate the visual quality of the volumetric representations using a 5-point Likert scale, while Prins et al. (2018) used a similar scale to evaluate the overall visual quality. Similarly, Gunkel et al. (2018) asked the participants of their study to rate the visual and audio quality using a 7-point Likert scale. As shown in **Table 3**, other low-level features have been also addressed in user studies, such as usability (typically evaluated with the System Usability Scale (SUS) by Brooke (1996) as in Bai et al. (2020); Teo et al. (2020)), exploration (e.g., using objective measurements of the head movements of the users while performing the tasks, as in Kasahara et al. (2017)), and responsiveness (Brunnström et al., 2020).

Finally, semi-structured interviews have been also used to evaluate diverse high and low-level factors, as done by Orts-Escolano et al. (2016) to assess presence, interactivity, exploration, believability, visual quality, and latency, and by Anjos et al. (2019) to assess task performance.

## 3.3 Conditions

While few user studies are limited to the evaluation of the features to assess the overall system performance, such as Prins et al. (2018); Gunkel et al. (2018); Anjos et al. (2019), most of them establish different experimental conditions for its validation. For example, some studies compare the performance of the analyzed system with previous technologies [e.g., traditional videoconferencing systems in Lawrence et al. (2021)] or with alternative immersive technologies and/or the face-to-face scenario in Li et al. (2019). Other works test different conditions based on certain technical aspects of the systems under study. For example, testing different displays has been used to evaluate the performance of different system archetypes, such as Social XR [e.g., HMD vs. 2D screen with game controller in Li et al. (2021), or VR vs. AR HMD in Orts-Escolano et al. (2016)], immersive telepresence [e.g., HMD vs. tablet (Zhang et al., 2020)], and face-to-face window [e.g., big vs. small displays in Lawrence et al. (2021)]. For remote assistance systems, user studies typically address the comparison of different visualization techniques of pointers and cues (e.g., cursors, head pointers, eye pointers, hand gestures, etc.) to support the indications from the remote user (Teo et al., 2019a; Bai et al., 2020; Wang et al., 2020), as well as different ways of interaction between the host and the visitor (Young et al., 2019). Also, the assessment of technical aspects related to the acquisition of the video from the host and the rendering for the visitor is relevant in these environments, such as the host's video stabilization in Kasahara et al. (2017) and visitor's view dependency on the host movements in Lee et al. (2018); Teo et al. (2020); Piumsomboon et al. (2019). Other conditions that influence higher-level features of the user experience can provide also insights of the performance of the systems, such as the use of different types of embodiment in Shared VR [e.g., with or without self-avatar in Pan and Steed (2017)] or the positioning of the visitor representation (e.g., video, volumetric representation, etc.) in immersive telepresence and Social XR systems (Lawrence et al., 2018;

Zhang et al., 2020; Piumsomboon et al., 2018, 2019). Finally, systems that do not consider communication also study technical aspects related to the interaction of the users with the systems, such as latency (Brunnström et al., 2020) or the use of controllers (Villegas et al., 2020).

## 3.4 Tasks

Taking into account the communication tasks considered in the reviewed user studies, we have classified them in three categories:

- **Deliberation:** Conversations between peers, normally oriented to achieve a common goal.
- **Exploration:** Exploration of the environment and identification of objects following indications.
- **Manipulation:** Interaction with system elements and manipulation of physical objects (e.g., lego blocks).

It is worth noting that all the tasks are communication tasks, and therefore they require a conversation between the users. The main difference is the nature of the conversation: in deliberation tasks the conversation *is the task*, in exploration tasks the conversation is used to discuss about the immersive environment, and in manipulation tasks the conversation is used to cooperate (or compete) in the execution of the task (e.g., provide instructions), which usually involves object interaction.

Various tasks related to deliberation are used in several studies to test the communication capabilities of almost all the system archetypes, especially those involving the Face element (**Table 1**). For instance, in Social XR/VR environments where shared experiences are provided to the users, such as watching movies together (Li et al., 2021), sharing photos (Li et al., 2019), and playing games (Gunkel et al., 2018), the deliberation task focused on commenting on the shared content. This was also studied in face-to-face window systems as in Kim S. et al. (2019). Other studies that do not use shared contents emphasize the deliberation proposing tasks such as negotiation (Lawrence et al., 2018), playing games (e.g., riddles in Anjos et al., 2019, tell a lie in Orts-Escolano et al., 2016, etc.) or using semi-structured interviews (Lawrence et al., 2021). Finally, in remote assistance and telepresence systems, which entail remote collaboration, identifying objects of interest can serve to test the communication capabilities in scenarios (Piumsomboon et al., 2019).

Exploration tasks are used in several user studies to test systems involving the Visit element, in which the visitor sees the environment of the host. For instance, identifying and locating objects (e.g., Lego blocks) in the task space is a widely used task in remote assistance scenarios (Kasahara et al., 2017; Lee et al., 2018; Teo et al., 2019a, 2020; Bai et al., 2020; Wang et al., 2020). In this type of systems, as well as in immersive telepresence and digital twin systems, the proposed tasks can also involve the exploration of the remote environment (Young et al., 2019; Piumsomboon et al., 2018, 2019). Also, the usefulness of visual cues, such as pointers (Bai et al., 2020; Wang et al., 2020) and eye-gaze (Zhang et al., 2020) on the remote collaboration can be assessed with exploration tasks. In this sense, Social XR systems can also include tasks related to the exploration of the shared environment as studied by Li et al. (2021), where the users were asked to follow the characters of the VR movie through the scene.

Finally, a variety of manipulation tasks have been used in different studies involving the Meet and Move elements. Several user studies test the capabilities of remote assistance and immersive telepresence systems proposing tasks where the host manipulates physical objects [e.g., Lego blocks in Wang et al. (2020); Bai et al. (2020)] or places them in the task environment (Lee et al., 2018; Teo et al., 2019a; Piumsomboon et al., 2018, 2019) following the indications of the visitor. These tasks have been also exploited in Social XR systems, such as in Lawrence et al. (2018) and Orts-Escolano et al. (2016). Nevertheless, Social XR/VR systems offer the possibility to test the interact with the environment, such as in Li et al. (2021), where the users could press buttons, or with virtual objects, such as in Pan and Steed (2017) and Prins et al. (2018), where the users played together puzzles and pong games, respectively. In this sense, it is also worth mentioning the insights that can provide user studies focused on analyzing the interaction of the users with the systems without accounting for communication, such as the one carried out by Brunnström et al. (2020) with a crane-control system and by Villegas et al. (2020) considering a scape room scenario.

## 4 DISCUSSION

There have been numerous attempts of designing, building, and testing immersive communication systems. Even though the technology has experimented a profound evolution over the last two decades, there is still no standard way to approach the problem, not even a *de facto* one. As a consequence, a diversity of systems has been designed and proposed in the literature, each one typically proposing its own evaluation methodology. At a first glance, this makes it difficult to propose a common analysis or evaluation framework for this variety.

However, our analysis shows that this apparent diversity can be structured around a few categories:

- There are not many different types of systems. Most of the systems can be classified according to whether and how they implement four basic communication features: *Face* (visual communication), *Visit* (remote presence), *Meet* (shared immersion), and *Move* (embodied interaction). As a result, most systems map into one of the archetypes identified in **Section 2.2**.
- Most evaluation protocols have a similar structure: they assess high-level or low-level QoE *features* under different *conditions* (different systems or different system configurations), using a *communication task* for the evaluation.
- There are not many types of tasks used for this evaluation. Most of them can be described as deliberation, exploration, or manipulation.

We will now discuss some of these aspects in detail, so that we are able to create a list of good practices for QoE assessment of immersive communication systems.

## 4.1 Types of Systems and Their Relation With QoE Features

The taxonomy of immersive systems that we have presented in **Section 2** is based on four communication elements which are *perceptual*, i.e., they describe what the user *can see* (or do) in the system. As shown in **Section 2.3**, those elements are closely related to the technology components of the communication system. Besides, they have a similar structure as the main components of presence proposed by Biocca (1997), so we could hypothesize that *Face* or *Meet* property should have a positive impact in social presence, *Visit* in place presence, and *Move* in self-presence. Even though the hypothesis is probably correct for many existing systems, it must be used carefully: system factors can significantly influence sense of presence, but the relationship between system and presence is not trivial.

For instance, Li et al. (2019) showed that a Shared VR system (*Meet*) can increase social and spatial presence with respect to conventional videoconference. Kim S. et al. (2019) showed the positive effect of *Face* in emotion conveyance. Cortés et al. (2020) and Young et al. (2019) showed that having full real-time immersive view of the remote location (*Visit*) has a significant impact in spatial presence compared to partial or non-updated views. And Villegas et al. (2020) showed that embodied interaction (*Move*) increase the sense of embodiment, but also spatial presence.

However, other influencing factors are also relevant for sense of presence, such as the type of visual content (Baños et al., 2004; Orduna et al., 2022) or individual differences such as personality (Alsina-Jurnet and Gutiérrez-Maldonado, 2010) or spatial intelligence (Jurnet et al., 2005). In fact, sense of presence does not require immersive visual stimuli; it is possible to elicit presence even with voice communication (Lombard and Ditton, 1997).

In summary, all the systems under consideration can provide high-level QoE features: Elicit sense of presence, enhance immersion, or improve interpersonal communication and empathy. They all potentially suffer from mental load issues too. And in all of them it is possible to evaluate similar low-level features, such as task performance, audiovisual quality, or interactivity. However, the specific details of which questionnaires are more suitable, which presence factors are more relevant, or which system factors are more critical will depend on the specific system and on the designed task.

## 4.2 Evaluation Protocols to Assess High-Level and Low-Level QoE Features

The experiments described in **Section 3** have been designed in a way which is possible to assess the difference of some high-level QoE factors (response variables) when executing a task under different system conditions. An important implication is that, to elicit high-level features such as presence or mental load, each task needs to be several minutes long. This reduces the number of possible conditions that can be tested in each experiment.

Besides, it is expected that high-level features only experiment significant differences where the test conditions are sufficiently different among them. Consequently, in the described experiments, only a few conditions are tested, and they are one of this two cases:

- Comparison of different communication systems. E.g., comparing the immersive system under study with conventional videoconference.
- Comparison between distinct options of key system features shown in **Figure 3**. E.g., comparing a screen with an HMD, or different modes of interaction.

A drawback of this approach is that the same restrictions on the assessment methodology are applied to low-level features, which are known to respond to finer-grain variations of the system conditions, and which can be evaluated with much shorter stimuli. For instance, ITU-T P.919 recommends 10-second sequences to assess visual quality of 360° video.

The conventional solution for this problem is conducting targeted experiments to only assess low-level features with shorter stimuli. For instance, to evaluate the visual quality of 360° (Gutiérrez et al., 2021) or point clouds (Viola et al., 2022), or the effect of delay in task performance (Brunnström et al., 2020). As an alternative approach, Orduna et al. (2022) propose a methodology to combine frequent assessment of low-level features during the execution of the test with the evaluation of high-level features using post-experience questionnaires. In particular, visual quality is evaluated each 25 s using Single-Stimulus Discrete Quality Evaluation (SSDQE) (Gutiérrez et al., 2011), while a set of high-level features (spatial and social presence, empathy, memory) is evaluated after each 5-minute sequence.

The evaluation of low-level features has been widely covered by ITU-T Recommendations. ITU-T P.919 proposes scales for video quality evaluation (e.g., ACR, *Absolute Category Rating*) and simulator sickness, as well as a method for the analysis of head and eye tracking data based on the works of David et al. (2018) and Fremerey et al. (2018). ITU-T P.1305 proposes techniques to analyze the conversation structure, which is useful to assess the effect of latency in the communication. ITU-T P.1312 proposes a framework for task performance analysis.

## 4.3 Selection of Tasks for the Evaluation of Immersive Communication Systems

The design of the tasks is normally related to the fundamental communication elements which are provided by the system. Even though any system (including audioconference!) can, in principle, be used for deliberation, exploration and manipulation tasks, some tasks which are more appropriate for each type of system.

**Face**-enabled systems are usually evaluated using deliberation tasks, since face-to-face contact is supposed to enhance a conversation compared to less immersive setups, including audioconference. **Visit**-enabled systems allow the visitor to explore the environment of the host, and therefore exploration tasks are suitable for them. Both Face and Visit-enabled systems support manipulation tasks, normally in the form where one user (the host in the case of Visit systems) performs the task and the other provides support or instructions.

**Meet**-enabled systems rely mostly on cooperative (or competitive) manipulation tasks, where all the users manipulate the same (virtual) objects. Conversation and (virtual) exploration tasks are also useful. Finally, **Move** features are also tested with manipulation tasks, normally jointly with the appropriate communication function.

## 4.4 Good Practices

As it has been seen, there are significant similarities between immersive communication systems, as well as between the evaluation methodologies proposed to assess their QoE. Based on them it is possible to propose a set of best practices for the QoE assessment of immersive communication systems, even though a standardized methodology does not exist yet.

1. **Characterize the system under test** according to the taxonomy defined in **Section 2**: Which of the fundamental elements (Face, Visit, Meet, Move) are implemented in the system and which archetype can be applied to it.
2. **Identify the relevant QoE features** to measure. High-level features should reflect the expectation of the use case foreseen for the system (what it is going to be used for), while low-level factors should address the most relevant technical variables of the system itself.
3. **Determine the system factors** that are going to be tested: whether the system is being **compared with others** or **different configurations/conditions** of the same system are being tested. In the former case, verify whether **all the systems under consideration share the same fundamental elements** or not.
4. **Identity a set of tasks** which can react to the variation of the experimental condition and allow to evaluate the QoE features under consideration. As a minimum, consider using the tasks appropriate for the fundamental elements: deliberation (Face), exploration (Visit), manipulation of a shared object (Meet), manipulation of an interactive element (Move). Pay special care in the design of the tasks if the systems under test do not share the same fundamental elements, as there is a high risk that the task definition biases the results towards one specific system.
5. **Define the test session in detail.** Adjust the duration of the tasks to find the proper trade-off between having a sufficiently stable environment to allow for high-level QoE features to arise and allocating the maximum number of test conditions. Define the steps of the experiment and their timing, how the different tasks or subtask are executed, and how the different conditions are tested.
6. **Select the right assessment tool** for each QoE factor. Select high-level evaluation questionnaires according to the nature of the system and the task. For low-level measures, consider using performance evaluation and other non-intrusive strategies as much as possible.

## 5 CONCLUSION

Users of immersive communication technologies, such as extended reality (XR), can explore and experience the stimuli in a more interactive and personalized way than previous technologies (e.g., 2D video). Thus, considering the new technological challenges related to these systems and the new perceptual dimensions and interaction behaviors involved, a deep understanding of the users' Quality of Experience (QoE) is required to satisfy their demands and expectations, especially in what regards the quality of communication and interaction between people. It is therefore needed to establish some common metrics and methodologies that can be used to assess such QoE.

In the paper we provide an overview of existing immersive communication systems and related user studies. Our analysis shows that the apparent diversity of systems and methodologies can be grouped around a few categories.

We have identified four fundamental elements of immersive communication systems: Face, Visit, Meet, and Move. We have shown that no existing system fulfills all of them and the same time, and therefore we have created a taxonomy of systems according to which of them they focus on. Particularly with Face, Visit, and Meet elements, systems typically address only one or two of them. We have also found that the systems which address the same combination of these elements tend to be based o the same set of technological components, which has allowed us to define a set of immersive communication system archetypes.

We have also identified the commonalities of the assessment methodologies used in the literature when testing this type of systems. Most evaluation protocols are similar: the compare two similar systems, or a few configurations of the same system, using a few communication tasks: deliberation, exploration, or manipulation. We have also identified the most relevant QoE features addressed in the literature, both high-level user experience traits and low-level system features.

With this analysis, we have provided a set of simple good practices for the QoE evaluation of immersive communication systems. We expect our contribution to be relevant to the community of immersive communication as well as VR, AR, MR, and XR, so that other researchers can build upon our proposed analysis framework. We also intend to feed this contribution into relevant standardization communities, such as the Video Quality Experts Group (VQEG) or the ITU-T, to help in the design of the next generation of subjective assessment methodologies for immersive communication systems.

## AUTHOR CONTRIBUTIONS

PP was the main contributor to the conceptualization of the work, with the support from the rest of the authors. PP, EG-S, and JG worked on the investigation process and writing of the manuscript. All authors contributed to manuscript revision, read, and approved the submitted version.

## FUNDING

This work has been partially supported by project PID2020-115132RB (SARAOS) funded by MCIN/AEI/10.13039/

501100011033 of the Spanish Government. The work of JG was partially supported by a Juan de la Cierva fellowship (IJC2018-037816) of the Ministerio de Ciencia, Innovación y Universidades of the Spanish Government. This work has been partially funded by the European Union's Horizon 2020 research and innovation programe under grant agreement No 957216 (Next-GENeration IoT sOlutions for the Universal Supply chain – iNGENIOUS.

**Conflict of Interest:** PP and EG-S was employed by the company Nokia.

The remaining authors declare that the research was conducted in the absence of any commercial or financial relationships that could be construed as a potential conflict of interest.

**Publisher's Note:** All claims expressed in this article are solely those of the authors and do not necessarily represent those of their affiliated organizations, or those of the publisher, the editors and the reviewers. Any product that may be evaluated in this article, or claim that may be made by its manufacturer, is not guaranteed or endorsed by the publisher.